\documentclass[english]{cplslarge}%
\usepackage[numbers]{natbib}
\usepackage{color}
\pdfoutput=0
\usepackage[T1]{fontenc}
\usepackage[latin9]{inputenc}
\usepackage{amsmath}
\usepackage{babel}

\begin{document}
\mainmatter
\author[Bouchet, Nardini \&\ Tangarife]{f. bouchet, c. nardini and t. tangarife}
\chapter{Kinetic theory and quasilinear theories of jet dynamics}

Regretfully, Tom\'as Tangarife suddenly and unexpectedly passed away few months before completing the writing of this chapter. Most of the science discussed in this text has been developed by a long and patient work by the three authors, including Tom\'as PhD thesis. Freddy Bouchet and Cesare Nardini pay homage to Tom\'as unique friendship and passion for science, and would like to remember the quiet, intense, and enriching collaboration that led to these scientific results.

\section{Introduction}

Turbulence in planetary atmospheres leads very often to self organisation
of the largest scales of the flow and to jet formation,
as discussed in many chapters of this book. We summarise here
a theory for jet formation and maintenance in a regime where velocity
fluctuations around the base jet are very small compared to the zonal
jet velocity itself. This regime is frequently present in the atmosphere of outer planets, the most prominent
example being probably Jupiter's troposphere jets, see chapters 2.3.8 and 2.3.9 of this book. Such
jets are continuously dissipated and forced by weak non-zonal turbulent
motion either from the deep atmosphere or due to the
differential heating of the planet. 
The balance between forcing and dissipation is mediated by
the non-zonal turbulent flow:
eddy dynamics, strongly affected by the jets, leads to
momentum flux convergence (Reynolds' stress divergence) that balance dissipation. This balance determines the
jet velocity profile. Moreover, for this regime the zonal jet themselves
are quasi-stationary: they evolve over time scales much longer than
the typical time scale of the non-zonal structures, as exemplified
for instance by comparison between Cassini and Voyager data for Jupiter's
zonal jets.

In such a regime, it is justified to treat the non-zonal
part of the dynamics with a quasi-linear approximation: at leading
order, the dynamics of the non-zonal flow is described by the 
equation linearised close to the quasi-stationary zonal jets. Such
quasi-linear approaches have been commonly studied for decades in
many theoretical discussions of geostrophic turbulence. Specifically
for the problem of jet formation, such a quasi-linear approach is
at the core of Stochastic Structural Stability Theory (S3T) first
proposed by Farrell, Ioannou \cite{bakasioannou2013,farrellioannou2007,farrellioannou2003},
for quasi-geostrophic turbulence, and discussed in section 5.2.2 of this book.
More recently, an interpretation in terms of a second order closure
(CE2) has also been given \cite{marston2011,marston2010,tobiasdagonetal2011,tobias2013}
(see section 5.1.2 of this book). All these different forms of quasi-linear
approximations have been extensively studied numerically, both using stochastic and deterministic forcings \cite{delsolefarrell1996}.
Very interesting empirical studies (based on numerical simulations)
have been performed recently in order to study the validity of this
type of approximation \cite{marston2011,marstonconoveretal2008,gormanschneider2007,tobias2013},
using the barotropic equations as well as more complex dynamics. The S3T
equations have also been used to study theoretically
the transition from a turbulence without a coherent structure to a
turbulence with zonal jets \cite{bakasioannou2013,parkerkrommes2013,srinivasanyoung2011}
(see sections 5.2.3, 5.2.4 of this book) and a generalisation aimed at studying the emergence of non-zonal structures is also discussed in section 5.2.5. These results are probably very close to approaches using Rapid Distortion Theory, or WKB Rapid Distortion Theory \cite{nazarenko2000,nazarenkokevlahanetal1999,nazarenkokevlahanetal2000}. We also observe that such a quasi-linear approach is classical in many other problems in theoretical physics for more than half a century, as it is for instance
at the core of the kinetic theory of plasmas and astrophysical systems (see for example
the derivation of Lenard-Balescu and similar kinetic equations in \cite{binneytremaine1987,bouchetguptaetal2009,landaulifshitz1996,nicholson1991,nardini2012kinetic,nardini2012kinetic2}). It is also a classical approach in fluid mechanics as it has been the base of the kinetic theory of point vortices and of two dimensional turbulence when dominated by large scale flows, for more than three decades. It is extremely useful to understand that all these physical problems fall into the same class of theoretical problems: the same tools may be developed and very interesting analogies may emerge. For this reason we refer to the quasilinear theory of zonal jet dynamics as an example of a kinetic theory.\\

The aim of this chapter is to discuss the theoretical aspects of such
a quasilinear description of statistically stationary jets. The basic
questions are: When is such an approach expected to be valid? Why? What are the limitations and the expected errors using such an
approximation? Should the deterministic S3T equations be corrected
by stochastic terms? Does such an approach describe only average states
or can it describe also fluctuations of the jet velocity profile? In situations where multiple attractors exist, like for instance for Jupiter's zonal jets, is it possible to compute transition rates between attractors from a quasilinear theory?

In order to address these issues, we study the jet formation problem
in the simplest possible theoretical framework: the two-dimensional
equations for a barotropic flow with a beta effect. These
equations, also called the barotropic quasi-geostrophic equations,
are the simplest relevant ones for the understanding of large scale planetary
flows \cite{pedlosky1982}. The theoretical approach summarized in this chapter could be in principle extended to the equivalent barotropic
quasi-geostrophic model (also called the Charney--Hasegawa--Mima equation),
to the multi-layer quasi-geostrophic models or to quasi-geostrophic
models for continuously stratified fluids \cite{pedlosky1982}, even
if the dynamics in those model is obviously of a different nature
as no baroclinic effects are modeled in the barotropic equations.

In statistical physics, kinetic theories are always associated with an asymptotic
expansion with respect to a small parameter. Our first message is that for turbulent barotropic flows on a beta plane such a non dimensional  parameter can be clearly identified \cite{bouchetsimonnet2008,bouchetvenaille2012}. It is denoted $\alpha$, and represents the ratio between i) an inertial time scale for the advection of small non-zonal eddies by the zonal jet and ii) the forcing time scale or equivalently the dissipation time scale (the spin-up or spin-down time scale, needed
to reach a statistically stationary energy balance). This will clearly answer our first question: a quasi-linear approach will be expected to be valid when this parameter is small. This is discussed in section \ref{sec:inertial-limit}.

In section \ref{sec:inertial-limit}, we present the barotropic model and discuss the range of parameters that leads to the formation of zonal jets. We also present the equation that describes the effective dynamics of zonal jets. The theoretical derivation of this equation is briefly presented in section \ref{sec:kinetic-theory}. This technical section can be entirely skipped at first reading. We then present the inviscid damping mechanism of the non-zonal eddies in section \ref{sec:Consistency}, answering the question: Why should a quasilinear approach be expected to be valid. Inviscid damping also allows to show that the time scale separation assumed to derive the effective equation of the slow zonal jet evolution is actually a self-consistent hypothesis. In section \ref{parameters}, we discuss comparison with numerical experiments, where we stress the expected errors using a quasilinear approximation. In section \ref{sec:bistability}, we discuss the fluctuations of the slow zonal jet dynamics. First we explain that a white in time noise can be easily added in order to describe Gaussian fluctuations. Then we explain how quasilinear approach can also be generalized in order to predict the large fluctuations that drive the dynamics from an attractor with a given number of zonal jets to a new attractor with either more or less zonal jets, as was observed on Jupiter in the past.

\section{The inertial limit and the effective slow jet dynamics}\label{sec:inertial-limit}
\subsection{Non-dimensional parameters and the inertial limit}
We study the formation of coherent structures in the barotropic equation
on a beta-plane, in a doubly periodic domain $\mathcal{D}=[0,2\pi Ll_{x})\times[0,2\pi L)$,
\begin{equation}
\partial_t q +\mathbf{v}\cdot\mathbf{\nabla}q=-\lambda\omega-\nu_{n,d}\left(-\Delta\right)^{n}\omega+\sqrt{\sigma}\eta,\label{eq:barotropic-d}
\end{equation}
where $\mathbf{v}=\mathbf{e}_{z}\times\mathbf{\nabla}\psi$ is the non-divergent velocity; $\omega=\Delta\psi$, $q=\omega+\beta_d y$, and $\psi$ are the vorticity, the potential vorticity and the stream function, respectively. $\lambda$ is a linear friction coefficient, $\nu_{n,d}$ is
a (hyper-)viscosity coefficient, and $\beta_d$ is the mean gradient
of potential vorticity. $\eta$
is a white in time Gaussian random noise, with spatial correlations
\[
\mathbf{E}\left[\eta({\bf r}_{1},t_{1})\eta({\bf r}_{2},t_{2})\right]=C({\bf r}_{1}-{\bf r}_{2})\delta(t_{1}-t_{2})
\]
that parametrize the forces (physically due, for example, to the effect of baroclinic instabilities or convection).
The correlation function $C$ is assumed to be normalised such that
$\sigma$ represents the average energy injection rate, so that the average
energy injection rate per unit of area (or equivalently per unit of mass taking into account density and the layer thickness) is $\epsilon=\sigma/4\pi^{2}L^{2}l_{x}$.\\
For atmospheric flows, viscosity is often negligible in the global energy balance and this is the regime that we will study in the following. Then
the main energy dissipation mechanism is linear friction. The evolution of the average energy (averaged over the noise realisations) $E$ is given by
\[
\frac{dE}{dt}=-2\lambda E+\sigma.
\]
In a stationary state we have $E=E_{stat}=\sigma/2\lambda$, expressing the balance between forces and dissipation. This relation gives the typical
velocity associated with the coherent structure $U\sim\sqrt{E_{stat}}/L\sim\sqrt{\epsilon/2\lambda}$. As will be clear in the following, we expect the non-zonal velocity perturbation to follow an inviscid relaxation, on a typical time scale proportional to the inverse of the shear rate.

For small values of $\beta_d$, it is expected that the structure is a jet at the largest scale of the box, so that a typical vorticity or shear is $s=U/L$
corresponding to a time $\tau=L/U$. It is then natural to define a non-dimensional parameter $\alpha$ as the ratio of the shear time scale over the dissipative time scale $1/\lambda$,
\[
\alpha=\lambda\tau=L\sqrt{\frac{2\lambda^{3}}{\epsilon}}.
\]
When $\alpha$ is small, there is a time scale separation between the relaxation time of the non-zonal perturbations  and the evolution of zonal jets. It is thus natural to derive an effective theory of the slow evolution of zonal jets using a small $\alpha$ expansion.

We write the non-dimensional barotropic equation using the box size $L$ as a length unit and the inverse of a typical shear $\tau=L/U$ as a time unit. We thus obtain (with a slight abuse of notation, due to the fact that we use the same symbols for the non-dimensional fields):
\begin{equation}
 \partial_t q +\mathbf{v}\cdot\mathbf{\nabla}q=-\alpha\omega-\nu_{n}\left(-\Delta\right)^{n}\omega+\sqrt{2\alpha}\eta,\label{eq:barotropic}
\end{equation}
with $q=\omega + \beta y$, where, in terms of the dimensional parameters, we have $\nu_n=\nu_{n,d}\tau/L^{2n}$, $\beta=\beta_d L \tau$. Observe that the above equation is defined on a domain $\mathcal{D}=[0,2\pi l_{x})\times[0,2\pi )$ and the averaged stationary energy for $\nu_n\ll \alpha$ is of order one. In the following, we will consider the case of viscosity, $n=1$, and denote $\nu=\nu_1$, but all the results can be generalized to any type of hyper-viscosity.

We observe that when the beta effect is large enough, several jets develop. Many works in literature \cite{vallis2006} suggest that the largest relevant scale of the flow is then given by the Rhines scale
\[L_{R}=\left(U/\beta_d \right)^{1/2}=\left(\epsilon/\beta_d^{2}\lambda\right)^{1/4}.\]
Such an estimate is actually relevant for $L_R\leq L$. In this regime, the Rhines scale gives actually the order of magnitude of the meridional jet width. Then a typical shear rate is $s=U/L_R$ corresponding to a time $\tau_R=L_R/U$. Then the ratio of the shear and dissipation time scales is
\[
\alpha_R=\lambda\tau_R=L_R\sqrt{\frac{2\lambda^{3}}{\epsilon}}.
\]
$\alpha_R$ is then be the natural expansion parameter in order to obtain an effective theory of the slow evolution of zonal jets.  We recognise that $\alpha_R  \propto\left(R_{\beta_d}\right)^{-5}$ where $R_{\beta_d}=\beta_d^{1/10}\epsilon^{1/20}\lambda^{-1/4}$ is the zonostrophy index used in many references. We thus conclude that when $L_R\leq L$, the kinetic theory regime, the regime in which the quasilinear approach is expected to be valid, is the regime when $\alpha_R\ll1$, or equivalently when $R_{\beta_d}\gg1$. We note that it is indeed observed in numerical simulations \cite{danilovgurarie2004,galperinsukorianskyetal2010} that the perturbations around zonal jets decrease when $R_{\beta_d}$ increases, as expected from this discussion. 

In the following, for simplicity we consider only the non-dimensional equations obtained using $\tau$ as the time unit, the natural one for $L_R\geq L$. Developing the theory for the non-dimensional equations obtained using $\tau_R$ as time unit, the natural one for $L_R\leq L$, would however be very similar. Moreover we note that $\alpha_R\leq\alpha$ when $L_R\leq L$. Thus, the hypothesis $\alpha\ll1$ made in the following actually implies $\alpha_R\ll1$. In section \ref{sec:numerics}, from some specific numerical simulations results, we discuss how small should actually be $\alpha$ or $\alpha_R$ in order to be in the kinetic theory regime.

\subsection{The effective slow zonal jet dynamics}\label{summary}
As eddies are weak with respect to the zonal jet in many physical situations, our main goal is to describe the effective evolution of the zonal degrees of freedom integrating out the effect of the eddies. As explained in the previous section, and discussed more precisely in section \ref{sec:Consistency}, when $\alpha \ll 1$ the eddies relax to a stationary state on a time scale much shorter than the time for the evolution of the jet. For this reason we investigate the range of parameters $\nu\ll\alpha\ll1$, called inertial limit.  The mathematical approach is called stochastic averaging, or adiabatic treatment \cite{gardiner1994}. In this section we describe the main result, the kinetic equation (\ref{eq:Stochastique-Lent-U-1}), and its consequences for the dynamics of slow jets. In the next sections, we will describe the derivation of the kinetic equation.

To extract the jet degrees of freedom out of the velocity field $\mathbf{v}$, we introduce the zonal average 
\begin{equation}
U(y)\equiv\left\langle v^{(x)}(x,y) \right\rangle =\frac{1}{2\pi l_x}\int \mathrm{d}x\,v^{(x)}(x,y)\,;
\end{equation}
the jet velocity profile  that we want to describe  is thus $(U(y),0)$.  The zonal part of the vorticity field will be denoted by $q_z=\langle q \rangle$.
The non-zonal part of the velocity will be denoted by a subscript $m$: 
\begin{equation}
\sqrt{\alpha}\mathbf{v}_m=\sqrt{\alpha}\left(v_m^{(x)},v_m^{(y)}\right)=\mathbf{v}-(U,0)\,,\label{decomposition-velocity}
\end{equation}
and analogous expressions for vorticity and stream-function fields. We also define the zonal and non-zonal parts of the noise as $\eta = \eta_z +\eta_m$, and $\zeta_z$ the effect of $\eta_z$ on the zonal jet $U$, such that $\eta_z = -\partial_y\zeta_z $. Observe the presence of $\sqrt{\alpha}$ in the definition of the non-zonal fields, which express the fact that non-zonal fluctuations are weak with respect to the mean flow. This is equivalent to assume the presence of a time-scale separation. The fact that this choice is actually a consistent hypothesis is one of the main points of our work; it will be discussed all along the chapter.

Our main result can be described as follows: in the limit $\nu\ll\alpha \ll 1$, the dynamics of the zonal jet velocity profile $U$ is described by the kinetic equation
\begin{equation}
\frac{1}{\alpha}\frac{\partial U}{\partial t}=\mathbf{E}_{U}\left[\left\langle v_{m}^{(y)}\omega_{m}\right\rangle \right]-U+\frac{\nu}{\alpha}\frac{\partial^{2}U}{\partial y^{2}}+\sqrt{2}\zeta_z+\sqrt{\alpha}\xi[U]\,,\label{eq:Stochastique-Lent-U-1}
\end{equation}
where $\omega_m$ solves 
\begin{equation}
\partial_t \omega_m + L^0_U[\omega_m]=-\alpha \omega_m +\nu \Delta \omega_m + \sqrt{2}\eta_m\,,\label{linear_dynamics}
\end{equation}
where
\vspace{0.2cm}
\begin{itemize}
\item $L_U^0$ is the advection operator linearised around $U$; explicitly, we have
\begin{equation}
L^0_{U}\left[\omega_{m}\right]=U(y)\partial_x\omega_{m}+\left(\partial_y q_z\right)\partial_x\psi_{m}\,.\label{eq:Linearized-Dynamics}
\end{equation}
Observe that the eddies evolve according to the linearized advection operator because their amplitude is of order $\sqrt{\alpha}$ smaller then the mean flow.
\item $\mathbf{E}_{U}[\cdot]$ is the average of the quantity in brackets over the stationary measure of the equation (\ref{linear_dynamics}). Explicitly, we have
\begin{equation}
\mathbf{E}_{U}\left[f[\omega_m] \right]=\lim_{t\to\infty}\mathbf{E}_{m}[f[\omega_m]]\label{eq:EU-definition}
\end{equation}
for any functional $f$, where $\mathbf{E}_{m}$ is the average over realisations of the noise $\eta_m$. $\left\langle v_{m}^{(y)}\omega_{m}\right\rangle$ is the zonally averaged momentum flux convergence, then $\mathbf{E}_{U}\left[\left\langle v_{m}^{(y)}\omega_{m}\right\rangle \right]$, that may be called a Reynolds stress divergence, is the statistical average of the momentum flux convergence. We note that $\mathbf{E}_{U}\left[\left\langle v_{m}^{(y)}\omega_{m}\right\rangle \right]$ can be computed directly from the two points correlation function for the vorticity derived from Eq. (\ref{linear_dynamics}). \\
Clearly, the presence of a long-time limit in the averaging procedure of the above quantity is due to the fact that a time scale separation is present in the system: eddies evolve much faster (on a time scale of order one) than the zonal jet, which evolves only on a time scale of order $1/\alpha$.
\item $\xi[U]$ is a stochastic term, that depends on the velocity profile $U$. Its correlation function is denoted by
\begin{equation}
\mathbf{E}[\xi[U](y_1,t_1)\,\xi[U](y_2,t_2)]=\Xi_{NL}[U](y_{1},y_{2})\delta(t_1-t_2)\,.
\end{equation}
$\Xi_{NL}$ accounts for the effects of fluctuations of large but finite time averages of the momentum flux convergence $\left\langle v_{m}^{(y)}\omega_{m}\right\rangle$. The expression for $\Xi_{NL}$ can be derived using a Green-Kubo formula, that can be evaluated from the two point-two times vorticity correlation function, as explained in  \cite{bouchetnardinietal2013}.\\
\end{itemize}

Let us discuss the physical properties of each of the terms of the kinetic equation (\ref{eq:Stochastique-Lent-U-1}). First of all, no hidden $\alpha$ nor $\nu$ dependencies are present in the kinetic equation. That means that in the considered regime $\nu\ll\alpha\ll 1$, the stochastic term $\sqrt{\alpha}\xi[U]$ is negligible. At first order in our perturbative expansion, the kinetic equation reduces to
\begin{equation}
\frac{1}{\alpha}\frac{\partial U}{\partial t}=\mathbf{E}_{U}\left[\left\langle v_{m}^{(y)}\omega_{m}\right\rangle \right]-U+\sqrt{2}\zeta_z\,.\label{eq:Stochastique-Lent-U-first-order}
\end{equation}
The deterministic evolution of the zonal jet is dictated by the first two terms on the r.h.s. of eq. (\ref{eq:Stochastique-Lent-U-first-order}). The first one is the momentum flux convergence $v_{m}^{(y)}\omega_{m}$ averaged both on the zonal direction (the symbol $\langle \cdot \rangle$) and according to the average $\mathbf{E}_{U}$ described above. The second one, $-U$, is just the direct effect of linear friction on the jet profile. At this order, fluctuations of the zonal jet profile are only given by $\zeta_z$, expressing the direct effect of the forcing on the zonal jet. 

From eq. (\ref{eq:Stochastique-Lent-U-first-order}), it is evident that the deterministic evolution of zonal jet profile is very slow, on a time scale of order $1/\alpha$. We should however observe that a subtlety may arise and break this conclusion: it is not obvious that $\mathbf{E}_{U}\left[\left\langle v_{m}^{(y)}\omega_{m}\right\rangle \right]$ has a limit in the inertial limit $\alpha \rightarrow 0$. Indeed, a large time limit enters in the definition of $\mathbf{E}_U$, see eq. (\ref{eq:EU-definition}) and eddies evolve according to equation (\ref{linear_dynamics}) where no dissipation is present in the aforementioned limit.

It is actually true that the statistical average of the momentum flux convergence $\mathbf{E}_{U}\left[\left\langle v_{m}^{(y)}\omega_{m}\right\rangle \right]$ may diverge if no further hypothesis are assumed on the base flow $U$, and then the asymptotic expansion would break down and the validity of the kinetic equation (\ref{eq:Stochastique-Lent-U-first-order}) would be very unlikely. For example, this happens if $U$ has unstable or neutral modes. In section \ref{sec:Consistency}, we will explain the steps of the theoretical justification that the statistical average of the momentum flux convergence is finite if $U$ has no unstable nor neutral modes. As a consequence we expect that under the hypothesis that $U$ has no unstable nor neutral modes, the slow evolution of $U$ on a time scale of order $1/\alpha$ is actually described at leading order by deterministic part of the kinetic equation (\ref{eq:Stochastique-Lent-U-first-order}).

It is also important to observe that the statistical average of the momentum flux convergence is a functional of $U$. This means that, in general the kinetic equation (\ref{eq:Stochastique-Lent-U-first-order}) may admit more than one attractor for fixed values of the physical parameters. This will be of importance in section \ref{sec:bistability}.

Eq. (\ref{eq:Stochastique-Lent-U-first-order}) is very similar to equations already introduced in the literature on a phenomenological ground (S3T and CE2, see \cite{bakasioannou2013,srinivasanyoung2011,tobiasdagonetal2011} and the following chapters of this book), and should coincide in the inertial limit $\alpha \rightarrow 0$. Their precise relation is discussed in section \ref{parameters}.\\

At next order in the kinetic equation (\ref{eq:Stochastique-Lent-U-1}) the stochastic term $\xi$ arises. This subdominant correction has essential consequences, especially in the physically relevant case of no forcing acting at large scales: $\zeta_z=0$. Indeed, under such an assumption, the kinetic equation at leading order (\ref{eq:Stochastique-Lent-U-first-order}) gives a deterministic evolution that does not describe jet fluctuations. In the inertial limit $\alpha \rightarrow 0$, equation (\ref{eq:Stochastique-Lent-U-1}) with $\xi$, properly describes Gaussian fluctuations of the jet. Another situation of particular interest arises when the deterministic dynamics has more than one attractor.  The statistical properties of rare transitions between different attractors then requires to study the fluctuation of large but finite time averages of the momentum flux convergence. We note however that the Gaussian fluctuations described by $\xi$ may not be precise enough to describe the statistics of the rare transitions between attractors, and that one has then to study large deviations of finite time averages of the momentum flux convergence. This kind of question are one of the most interesting perspectives of our work, as further discussed in section \ref{sec:bistability}. In section \ref{sec:bistability} we also argue that this may be relevant for Jupiter's zonal jets.

\section{Stochastic averaging of the barotropic equations}\label{sec:kinetic-theory}
In this section we summarise the perturbative technique that permits to obtain formally, in the inertial limit $\nu\ll\alpha\ll1$, the kinetic equation (\ref{eq:Stochastique-Lent-U-1}) for the slow evolution of the zonal jet velocity profile $U$. This section follows a classical \cite{gardiner1994} but rather technical development for dynamical systems with a fast and slow time scale and can be entirely skipped at first reading. Moreover, not all the details will be given here and we address the interested reader to \cite{bouchetnardinietal2013}. In section \ref{sec:Consistency} we will go beyond this formal justification, by justifying the self-consistency of the hypothesis made by checking the orders of magnitude of the main terms in the asymptotic expansion.

\subsection{Decomposition into zonal flow and eddies}

Zonal jets are characterised by their velocity profile
${\bf v}({\bf r},t)=U(y,t){\bf e}_{x}$. From Eq. (\ref{eq:barotropic}), it is natural to assume that the turbulent
fluctuations are of order $\sqrt{\alpha}$. A major
part of this work, summarised in section \ref{sec:Consistency}, will
consist in proving that this assumption is self-consistent. Defining
the zonal projection $\left\langle .\right\rangle $ of a generic function $f$ as 
\[
\langle f\rangle (y)=\frac{1}{2\pi l_{x}}\int_{0}^{2\pi l_{x}}\mathrm{d}x\, f(\mathbf{r}),
\]
the zonal part of the potential velocity field will be denoted by
$U\equiv\left\langle {\bf v \cdot {\bf e}_{x}}\right\rangle $; the rescaled non-zonal
part of the flow ${\bf v}_{m}$ is then defined through the decomposition
\begin{equation}
\mathbf{v}({\bf r})=U(y)\mathbf{e}_{x}+\sqrt{\alpha}\mathbf{v}_{m}({\bf r}).\label{eq:scaling-fluctuations}
\end{equation}
Similarly, the potential vorticity will be denoted $q=q_{z}+\sqrt{\alpha}\omega_{m}$.

We now project the barotropic equation (\ref{eq:barotropic}) into
zonal 
\begin{equation}
\partial_t q_{z}=-\alpha\partial_y\left\langle v_{m}^{(y)}\omega_{m}\right\rangle -\alpha\omega_{z}+\nu\partial^{2}_y\omega_{z}+\sqrt{2\alpha}\eta_{z}\label{eq:Zonal-PV-Evolution}
\end{equation}
and non-zonal part 
\begin{equation}
\partial_t\omega_{m}+L_{U}\left[\omega_{m}\right]+\sqrt{\alpha}NL[\omega_{m}]=\sqrt{2}\eta_{m},\label{eq:Meridional-PV-Evolution}
\end{equation}
with the linear operator
\begin{equation}
L_{U}\left[\omega_{m}\right]=U(y)\partial_x\omega_{m}+q'_{z}(y)\partial_x\psi_{m}+\alpha\omega_{m}-\nu\Delta\omega_{m}\label{eq:Linearized-Dynamics-1}
\end{equation}
and the non-linear operator
\[NL[\omega_{m}]={\bf v}_{m}\cdot\nabla\omega_{m}-\left\langle {\bf v}_{m}\cdot\nabla\omega_{m}\right\rangle .\]
In the above equations, $\eta_{z}=\left\langle \eta\right\rangle $  (resp. $\eta_{m}=\eta-\left\langle \eta\right\rangle $)
is a white in time Gaussian noise with spatial correlation function $C_{z}=\left\langle C\right\rangle $ (resp. $C_{m}=C-\left\langle C\right\rangle $). Observe that the cross
correlation between $\eta_{z}$ and $\eta_{m}$ is exactly zero, due
to the translational invariance of $C$.

In the decomposed equations (\ref{eq:Zonal-PV-Evolution}), (\ref{eq:Meridional-PV-Evolution})
it is clear that the natural time-scale of evolution of $q_{z}$ is
of order $1/\alpha$ while the natural time-scale of evolution of
$\omega_{m}$ is of order 1. This is a direct consequence of our working ansatz that turbulent fluctuations are weak (\ref{eq:scaling-fluctuations}).

To proceed further, it is useful to work not at the level of the stochastic equations presented above but at the level of the associated functional Fokker-Planck equation. Thanks to the general theory of stochastic differential equations \cite{gardiner1994},  (\ref{eq:Zonal-PV-Evolution}) and (\ref{eq:Meridional-PV-Evolution})
are equivalent to the Fokker-Planck equation
\begin{equation}
\partial_t P=\mathcal{L}_{0}P+\sqrt{\alpha}\mathcal{L}_{n}P+\alpha\mathcal{L}_{z}P,\label{eq:Evolution-P}
\end{equation}
for the probability distribution function (PDF) $P[q_{z},\omega_{m}]$.
The distribution $P\left[q_{z},\omega_{m}\right]$ is a functional
of the two fields $q_{z}$ and $\omega_{m}$ and is a formal generalisation
of the probability distribution function for variables in finite dimensional
spaces.\\
We have divided the Fokker-Planck operator in three parts. The first one
\begin{eqnarray}
\mathcal{L}_{0}P\equiv &\int\mbox{d}\mathbf{r}_{1}\,\frac{\delta}{\delta\omega_{m}(\mathbf{r}_{1})}\bigg[L_{U}\left[\omega_{m}\right](\mathbf{\mathbf{r}}_{1})P\\
&+\int\mbox{d}\mathbf{r}_{2}\, C_{m}(\mathbf{r}_{1}-\mathbf{r}_{2})\frac{\delta P}{\delta\omega_{m}(\mathbf{r}_{2})}\bigg]
\end{eqnarray}
is the Fokker-Planck operator that
corresponds to the linearized dynamics (\ref{eq:Linearized-Dynamics-1})
close to the zonal flow $U$, forced by a Gaussian noise, white in
time and with spatial correlations $C_{m}$. This Fokker-Planck operator
acts on the non-zonal variables only and depends parametrically on
$U$. \\
At order $\sqrt{\alpha}$, the term
\[\mathcal{L}_{n}P\equiv\int\mbox{d}\mathbf{r}_{1}\,\frac{\delta}{\delta\omega_{m}(\mathbf{r}_{1})}\left[NL[\omega_m](\mathbf{r}_1)P\right]\]
contains the non-linear interactions between non-zonal
degrees of freedom.\\
At order $\alpha$, the term
\begin{eqnarray}
\nonumber\mathcal{L}_{z}P\equiv&\int\mbox{d}y_{1}\,\frac{\delta}{\delta q_{z}(y_{1})}\bigg[\left(\alpha\partial_y\left\langle v_{m}^{(y)}\omega_{m}\right\rangle +\alpha\omega_{z}-\nu\partial^{2}_y\omega_{z}\right)P\\
&\qquad +\int\mbox{d}y_{2}\, C_{z}(y_{1}-y_{2})\frac{\delta P}{\delta q_{z}(y_{2})}\bigg]
\end{eqnarray}
contains the terms that describe the coupling between
the zonal and non-zonal flow, the dynamics due to friction acting
on zonal scales and the zonal part of the stochastic forces.

Our goal now is to obtain a reduced Fokker-Planck equation that describes
only the slow evolution of the zonal jet $U$, using a perturbative expansion in the small parameter $\alpha\ll1$.

\subsection{The quasilinear eddy distribution}\label{sub:stationary-distribution-fast}
As previously stressed, in the limit $\alpha\ll 1$, there is a time scale separation between the evolution of $\omega_m$ and the evolution of $q_z$. It is thus simple to guess that, to develop the kinetic theory, we have first to determine the stationary distribution
of $\omega_m$, with $U$ held fixed.\\
Such stationary distribution is obtained by imposing $\mathcal{L}_{0}P=0$ where $U$ is considered as fixed. This stationary Fokker-Planck equation describes the statistically stationary state of the stochastic equation
\begin{equation}
\partial_t\omega_{m}+L_{U}\left[\omega_{m}\right]=\sqrt{2}\eta_{m},\label{eq:linearized-omegam}
\end{equation}
with the linear operator $L_U$ given by (\ref{eq:Linearized-Dynamics-1}). Equation (\ref{eq:linearized-omegam}) is a linear process (Ornstein-Uhlenbeck process), as a consequence its stationary measure is Gaussian for any initial state. Moreover, as $\mathbf{E}_m[\omega_{m}]=0$, the stationary distribution is completely characterised by the stationary two-points correlation function $g^{\infty}[q_{z}](\mathbf{r}_{1},\mbox{\ensuremath{\mathbf{r}}}_{2})=\lim_{t\to\infty}\mathbf{E}_m\left[\omega_{m}({\bf r}_{1},t)\omega_{m}({\bf r}_{2},t)\right]$, where $\mathbf{E}_m$ denotes the average over the realisations of
the noise $\eta_{m}$, for fixed $U$. \\
The two-points correlation function $g^{\infty}$ is the
stationary solution of the so-called Lyapunov equation, obtained from the It\={o} formula applied to (\ref{eq:linearized-omegam}), 
\begin{equation}
\partial_t g+L_{U}^{(1)}g+L_{U}^{(2)}g=2C_{m},\label{eq:equation-Lyapunov}
\end{equation}
where $L_{U}^{(i)}$ is the linearized operator $L_{U}$ defined in (\ref{eq:Linearized-Dynamics-1})
acting on the variable $\mathbf{r}_{i}$. From (\ref{eq:equation-Lyapunov}),
it is clear that $g^{\infty}$ depends on the base flow $U$
(or equivalently on $q_{z}$). As a consequence, all the quantities averaged
with the stationary distribution of (\ref{eq:linearized-omegam}), also depend parametrically on $q_{z}$. 

We
denote by 
\begin{equation}
G[q_{z},\omega_{m}] = \frac{1}{Z}e^{-\frac{1}{2}\int\mbox{d}\mathbf{r}_1\mbox{d}\mathbf{r}_2\,\omega_m(\mathbf{r}_1)\left(g^\infty [q_z]\right)^{-1}(\mathbf{r}_1,\mathbf{r}_2)\omega_m(\mathbf{r}_2)}\label{Gaussian-distribution}
\end{equation}
the Gaussian stationary distribution of (\ref{eq:linearized-omegam}) and by
\[\mathbf{E}_{U}[A]=\int\mathcal{D}[\omega_m] G[q_z,\omega_m]A[\omega_{m}]\]
the average of an observable $A[\omega_{m}]$ over the distribution
$G[q_{z},\omega_{m}]$.

The convergence of $g$ towards $g^\infty$ in the limit $t\to\infty$ implies the existence
of the stationary distribution $G[q_{z},\omega_{m}]$. It is thus
a crucial point of this theory and is related to the self-consistency
of the assumed scaling for the fluctuations (\ref{eq:scaling-fluctuations}).
This fundamental issue is discussed in section \ref{sec:Consistency}.

\subsection{Derivation of the slow dynamics of zonal jets \label{sub:Fokker-Planck-equation-for}}
To formalise the perturbative expansion of the Fokker-Planck equation (\ref{eq:Evolution-P}), we introduce the decomposition $P=P_{s}+P_{f}$
through the projection operator $\mathcal{P}$:
\[
P_s\equiv\mathcal{P}P\equiv G[q_{z},\omega_{m}]\int\mathcal{D}[\omega_{m}]\, P[q_{z},\omega_{m}],
\]
and $P_{f}\equiv(1-\mathcal{P})P$. The two PDF $P$ and $P_s$ differ because in the latter the turbulent fluctuations are relaxed to their stationary distribution
$G[q_{z},\omega_{m}]$. We also denote by 
\[R[q_{z}]=\int\mathcal{D}[\omega_{m}]P[q_{z},\omega_{m}]\] 
the marginal
distribution of the zonal jet, with the turbulence
averaged out.

The goal of the pertubative expansion (also called stochastic averaging) is to get a closed equation for
the evolution of $R$ from the complete Fokker-Planck equation
(\ref{eq:Evolution-P}). It follows classical methods \cite{gardiner1994},
and the explicit computations in this particular case are reported
in \cite{bouchetnardinietal2013}. The first step is to
apply the projections $\mathcal{P}$ and $1-\mathcal{P}$ on the Fokker-Planck
equation (\ref{eq:Evolution-P}):

\[\partial_{t}P_{s}=\alpha\mathcal{P}\mathcal{L}_{z}\left(P_{s}+P_{f}\right),\]

\begin{equation}\partial_{t}P_{f}=\mathcal{L}_{0}P_{f}+\left(\sqrt{\alpha}\mathcal{L}_{n}+\alpha(1-\mathcal{P})\mathcal{L}_{z}\right)(P_{s}+P_{f})\label{eq:evolution-Ps-Pf}.\end{equation}
In the above equations we have used $\mathcal{P}\mathcal{L}_{0}=\mathcal{L}_{0}\mathcal{P}=0$,
which is clear from the definition of $\mathcal{P}$, and $\mathcal{P}\mathcal{L}_{n}=0,$
due to the fact that $\mathcal{L}_{n}$ acts only on the
non-zonal degrees of freedom. As it has been anticipated by the notation, we clearly see in (\ref{eq:evolution-Ps-Pf})
the time-scale separation between the slow evolution of $P_{s}$ and the fast
evolution of $P_{f}$.\\
The equation on $P_{f}$ can be formally solved using Laplace transform,
and is then injected into the equation on $P_{s}$. This equation
is then expanded in powers of $\alpha$ to order $\alpha^{2}$.
Performing the inverse Laplace transform, we observe that the evolution
equation for $P_{s}$ contains memory terms. However, in the limit
$\alpha\ll1$, $P_{s}$ evolves very slowly and a Markovianization procedure can be employed.

At order $\alpha^{2}$, we obtain
\begin{eqnarray}
\frac{\partial P_{s}}{\partial t}=&\bigg\lbrace \alpha\mathcal{P}\mathcal{L}_{z}+\alpha^{3/2}\mathcal{P}\mathcal{L}_{z}\int_{0}^{\infty}\mbox{d}t'\,\mbox{e}^{t'\mathcal{L}_{0}}\mathcal{L}_{n}+\\
&\alpha^{2}\mathcal{P}\mathcal{L}_{z}\int_{0}^{\infty}\mbox{d}t'\,\mbox{e}^{t'\mathcal{L}_{0}}\bigg[(1-\mathcal{P})\mathcal{L}_{z}+\\
&\int_{0}^{\infty}\mbox{d}t''\,\mathcal{L}_{n}\mbox{e}^{t''\mathcal{L}_{0}}\mathcal{L}_{n}\bigg]\bigg\rbrace P_{s}(t)+\mathcal{O}\left(\alpha^{5/2}\right).\label{eq:Evolution-Ps-1}
\end{eqnarray}
The different terms above can then be computed explicitly \cite{bouchetnardinietal2013},
we discuss here the main aspects of this computation. The first term
in the right hand side of (\ref{eq:Evolution-Ps-1}) gives
the momentum flux convergence averaged over the stationary distribution $G[q_{z},\omega_{m}]$.
The next term vanishes exactly, because the non-linear interaction
term $NL[\omega_{m}]$ in $\mathcal{L}_{n}$ leads to the computation
of odd moments of the Gaussian distribution $G[q_{z},\omega_{m}]$.
At order $\alpha^{2}$, the first term produces a diffusion term, which corresponds to a (white in time) Gaussian noise,
and the last term represents a correction to the drift term due to
the non-linear interactions.

We do not enter in further details here; the interested reader can consult \cite{bouchetnardinietal2013}, in which the above computation is detailed. The result of this procedure is a Fokker-Planck equation for the slow evolution of the zonal jet PDF $R$
\begin{eqnarray}
\frac{1}{\alpha}\frac{\partial R}{\partial t}=& \int\mbox{d}y_{1}\,\frac{\delta}{\delta q_{z}(y_{1})}\left\{ \left[\frac{\partial F_{1}}{\partial y_{1}}+\omega_{z}(y_{1})-\frac{\nu}{\alpha}\frac{\partial^{2}\omega_{z}}{\partial y_{1}^{2}}\right]R[q_{z}]+\right.\nonumber\\
&\left.\int\mbox{d}y_{2}\,\frac{\delta}{\delta q_{z}(y_{2})}\left(C_{R}(y_{1},y_{2})R\left[q_{z}\right]\right)\right\} .\label{eq:Fokker-Planck-Zonal}
\end{eqnarray}
This Fokker-Planck equation can be recast in an equivalent stochastic differential equation for the potential vorticity profile $q_z(y,t)$
\begin{equation}
\frac{1}{\alpha}\frac{\partial q_z}{\partial t}=  -\frac{\partial F_{1}}{\partial y_{1}}-\omega_{z}(y_{1})+\frac{\nu}{\alpha}\frac{\partial^{2}\omega_{z}}{\partial y_{1}^{2}}+\eta[U],\label{eq:kinetic-qz}
\end{equation}
where $\eta[U]$ is a white in time Gaussian noise with spatial correlation $C_R$. In the above equations (\ref{eq:Fokker-Planck-Zonal},\ref{eq:kinetic-qz}), the drift
term is
\[
F_{1}=F\left[U\right]+\alpha \mathcal{M}[U],
\]
with
\[F\left[U\right]=\mathbf{E}_{U}\left[\left\langle v_{m}^{(y)}\omega_{m}\right\rangle \right]\,\]
and the explicit form of $\mathcal{M}$ can be found in \cite{bouchetnardinietal2013}. The diffusion coefficient is 
\[
C_{R}(y_{1},y_{2})=C_{z}(y_{1}-y_{2})+\alpha  \frac{\partial^2}{\partial y_1 \partial y_2}  \Xi_{NL}(y_{1},y_{2})\left[U\right],
\]
where we recall that $C_z$ is the zonal average of the correlation function $C$ of the original noise appearing in the barotropic equations (\ref{eq:barotropic}); the correlation function of the non linear part of the noise is given by
 \begin{eqnarray}\label{eq:CNL}
&\Xi_{NL}(y_{1},y_{2})\left[U\right] =\\& \int_{0}^{\infty}\mbox{d}t'\,\mathbf{E}_{U}\left[\left[\left\langle v_{m}^{(y)}\omega_{m}\right\rangle (y_{1},t')\left\langle v_{m}^{(y)}\omega_{m}\right\rangle (y_{2},0)\right]\right],\nonumber
 \end{eqnarray}
where $\mathbf{E}_{U}[[A(t)B(0)]]$ is the covariance of the observables $A[\omega_{m}]$ and $B[\omega_{m}]$, with respect to the Gaussian distribution (\ref{Gaussian-distribution}).\\

Equation (\ref{eq:kinetic-qz}) can be integrated to get the evolution equation of the jet velocity profile $U(y,t)$
\begin{equation}
\frac{1}{\alpha}\frac{\partial U}{\partial t}=   \mathbf{E}_{U}\left[\left\langle v_{m}^{(y)}\omega_{m}\right\rangle \right]+\alpha \mathcal{M}[U]-U+\frac{\nu}{\alpha}\frac{\partial^{2}U}{\partial y_{1}^{2}}+\sqrt{2}\zeta_z+\sqrt{\alpha}\xi,\label{eq:kinetic-U}
\end{equation}
where $\xi$ is a white in time Gaussian noise with spatial correlation
\begin{equation}
C_U(y_1,y_2)=\Xi_{NL}(y_{1},y_{2})\left[U\right]   
\end{equation}
with $\Xi_{NL}$ given in (\ref{eq:CNL}).

At second order in the kinetic equation (\ref{eq:kinetic-U}), there are two terms: a deterministic one, $\mathcal{M}$, and a stochastic one, $\xi$: both of them are functionals of $U$. We expect the effect of $\alpha\mathcal{M}$ to be only a small correction to the jet relaxation when $\alpha$ is small. We can thus neglect $\mathcal{M}$, and we obtain (\ref{eq:Stochastique-Lent-U-1}).

\section{Inviscid damping and consistency of the asymptotic expansion\label{sec:Consistency}}

The kinetic equation (\ref{eq:Stochastique-Lent-U-1}) that describes the slow dynamics of zonal jets involves the stationary average  $\mathbf{E}_U$ of the momentum flux convergence (the Reynolds stress divergence) $\mathbf{E}_{U}\left[\left\langle v_{m}^{(y)}\omega_{m}\right\rangle \right]$. It is not obvious that this average has a finite limit in the inertial limit $\alpha \rightarrow 0$. Indeed, the eddy dynamics (\ref{linear_dynamics}) is forced but not dissipated in this limit. If the limit of $\mathbf{E}_{U}\left[\left\langle v_{m}^{(y)}\omega_{m}\right\rangle \right]$ would not be finite, then the asymptotic expansion may break down and the kinetic equation or the related S3T dynamics would probably not be valid. The ergodicity of the momentum flux convergence (does the time averages of the momentum flux convergence converge to its statistical average?) is a also a necessary requirement for the theory to make sense. We now consider these two essential questions.  

As the dissipation on the eddy equation (\ref{linear_dynamics}) vanishes in the inertial limit, in order to have finite large time limits and ergodicity, we have to rely on an inviscid damping mechanism. In the case of the linearized 2D Euler equation such a mechanism is known as the Orr mechanism \cite{orr1907}. In this section we first recall classical results about the Orr mechanism for the two-dimensional Euler equation ($\beta=0$) \cite{orr1907}. We also discuss their generalisation to any jet profile \cite{bouchetmorita2010}, holding when the base flow has no modes neither unstable nor neutral.
Based on the Orr mechanism, we show that the statistical average of the momentum flux convergence $\mathbf{E}_{U}\left[\left\langle v_{m}^{(y)}\omega_{m}\right\rangle \right]$ has a finite limit in the inertial limit. Thus our kinetic equation is well defined at order $\alpha$ and the hypothesis of time-scale separation is self consistent, at least as far as the quasilinear momentum flux convergence is concerned. Finally, we consider the generalisation of these results to $\beta\neq 0$ and to the case when the base flow has neutral modes.

\subsection{Balance between dissipation and forcing}
It will be useful in the following discussion to have in mind a very simple example of the balance between dissipation and forcing in a stochastic dynamics. We consider the one degree of freedom Ornstein--Uhlenbeck process
\begin{equation}
\label{eq:q}
\frac{{\rm d} q}{{\rm d} t} = -\alpha q + \sqrt{\sigma}\eta(t)\,,
\end{equation}
where $\eta$ is a white in time Gaussian noise, $\alpha, \sigma >0$, and with initial condition $q(0)=0$. We investigate the large-time limit of the variance of $q$. Integrating equation (\ref{eq:q})
\begin{equation}
 q(t) = \sqrt{\sigma}\int_0^t \mbox{e}^{-\alpha (t-u)}\eta(u)\mbox{d}u\,,
\end{equation}
we get
\begin{equation}
 \mathbf{E}\left[q(t)^2\right] = \sigma\int_0^t \left[\mbox{e}^{-\alpha u}\right]^2\mbox{d}u\,,\label{autocorrelation_1D}
\end{equation}
where $\mathbf{E}$ denotes the average with respect to realisation of the noise $\eta$.

From this simple analysis, we can conclude that the convergence of the variance when $t\to\infty$ depends on the value of the friction coefficient $\alpha$. Indeed, if $\alpha >0$, the auto-correlation function converges to the finite value $\sigma/2\alpha$, while for $\alpha = 0$, the variance function diverges as $\sigma t$.

Observe that in equation (\ref{autocorrelation_1D}), the variance is expressed from the solution $\tilde{q}(u)= \mbox{e}^{-\alpha u}$ of the deterministic equation $\partial_t \tilde{q} = -\alpha \tilde{q}$ with initial condition $\tilde{q}(0)=1$. We thus conclude that the convergence of the variance depends on the large-time behaviour of the associated deterministic linear evolution, and particularly on the damping mechanism in this deterministic dynamics.\\

This discussion is very general, and an expression similar to (\ref{autocorrelation_1D}) can be obtained for any Ornstein-Uhlenbeck process \cite{gardiner1994}. The computation of auto-correlation functions can be discussed similarly to the computation of the variance. We thus understand that in the problem we are interested in, the convergence of the averaged momentum flux convergence $\mathbf{E}_{U}[\langle v_{m}^{(y)}\omega_{m}\rangle ]$ will depend on the large-time behaviour of the deterministic linear equation
\begin{equation}
\partial_t \tilde{\omega}_{m} + L^0_{U}[\tilde{\omega}_{m}]=-\alpha \tilde{\omega}_{m} +\nu \Delta \tilde{\omega}_{m}\,.\label{linear_dynamics-deterministic}
\end{equation}
For finite values of $\alpha$ and $\nu$, and for a stable linear operator $L^0_{U}$, the linear friction and viscosity are the main damping mechanisms. Then, the vorticity auto-correlation function, and the average momentum flux convergence do exist. Moreover, we are interested in the particular limit where $\nu \ll \alpha \ll 1$ and, for the self-consistency of the expansion we need a convergence rate independent of the values of $\nu$ and  $\alpha$. Then we need to rely on another damping mechanism, through the linear operator $L_U^0$.  For the linearized Euler equation, such inviscid damping mechanisms are known as the Orr mechanism and the depletion of vorticity at the stationary streamlines \cite{bouchetmorita2010}. These mechanisms are summarised in the following section.

Moreover, we can see directly from (\ref{autocorrelation_1D}) that if the linear operator is unstable, the deterministic evolution diverges exponentially, so the auto-correlation function also diverges. The same way, we see that if the linear operator has neutral modes, the auto-correlation function will diverge linearly in time. It is thus essential for the self-consistency of the expansion to assume that the base flow $U$ has no normal modes at all. This is possible for a linear operator acting in an infinite-dimensional space, such as $L_U^0$. Actually in many jets, the dynamics is known to actually expel neutral modes from the spectrum \cite{kasahara1980}. We discuss further this hypothesis in the following paragraphs.

\subsection{Orr mechanism and depletion of vorticity at the stationary streamlines}

\begin{figure}
\figurebox{}{15pc}{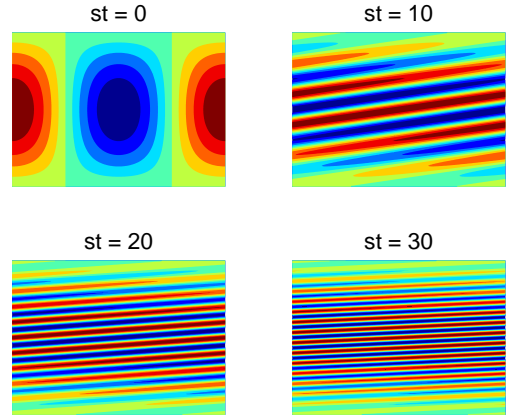}
\caption{Evolution of the perturbation vorticity, advected by the constant shear base flow $U(y)=sy$.\label{figure-orr}}
\end{figure}

We consider here the linear deterministic equation (\ref{linear_dynamics-deterministic}) with $\beta=0$, and with no viscosity or linear friction, $\alpha=\nu=0$. The phenomenology is the following: while the vorticity shows filaments at finer and finer scales when
time increases, non-local averages of the vorticity (such as the one leading to the computation of the stream-function or the velocity) converge to zero in the long-time limit. As an example, the filamentation can be seen in figure \ref{figure-orr}, for the vorticity field advected by a constant shear flow $U(y)=sy$. This filamentation and the related relaxation mechanism with no dissipation for the velocity and stream function is very general for advection equations and it has an analog in plasma physics in the context of the Vlasov
equation, where it is called Landau damping \cite{nicholson1991}.

In order to be more precise, we consider the deterministic linear dynamics $\partial_{t}\tilde{\omega}_m+L_{U}^{0}[\tilde{\omega}_m]=0$
with initial condition $\mbox{e}^{ikx}f(y)$. As explained at the end of the previous paragraph, it is natural to assume that the linear operator $L_{U}^{0}$ has no normal modes. With this hypothesis, it can be shown \cite{bouchetmorita2010} that the solution is of the
form $\tilde{\omega}_m(x,y,t)=\mbox{e}^{ikx}\tilde{\omega}_{k}(y,t)$ with, for $t$ going to infinity,
\begin{equation}
\tilde{\omega}_{k}(y,t)\sim\tilde{\omega}_{k}^{\infty}(y)\mbox{e}^{-ikU(y)t}\,.\label{eq:orr-mechanism-voricity}
\end{equation}
We thus see that the vorticity oscillates on a finer and finer scale
as time goes on. By contrast to the behaviour of the vorticity,
any spatial integral of the vorticity decays
to zero. For instance, the results for the $x$ and $y$ components
of the velocity and for the stream function are: 
\begin{equation}
\tilde{v}_{k}^{(x)}(y,t)\sim\frac{\tilde{\omega}_{k}^{\infty}(y)}{ikU'(y)}\frac{\mbox{e}^{-ikU(y)t}}{t},\label{eq:orr-mechanism-velocity-x}
\end{equation}
 
\begin{equation}
\tilde{v}_{k}^{(y)}(y,t)\sim\frac{\tilde{\omega}_{k}^{\infty}(y)}{ik(U'(y))^{2}}\frac{\mbox{e}^{-ikU(y)t}}{t^{2}},\label{eq:orr-mechanism-velocity-y}
\end{equation}
and 
\begin{equation}
\tilde{\psi}_{k}(y,t)\sim\frac{\tilde{\omega}_{k}^{\infty}(y)}{(ikU'(y))^{2}}\frac{\mbox{e}^{-ikU(y)t}}{t^{2}}\,.\label{eq:orr-mechanism-stream}
\end{equation}
In all the above formulas, higher order corrections are present and
decay with higher powers in $1/t$. From these expressions, it is clear that the local shear $U'(y)$ acts as an effective damping mechanism. This is the so-called Orr mechanism.

At this stage, a natural question is: what happens when the local shear vanishes? Indeed, a jet profile necessarily presents extrema of the velocity, at points $y_0$ such that $U'(y_0)=0$. Such points are called stationary points of the zonal jet profile. It can be shown that at the stationary points, the perturbation vorticity also decays for large times: $\tilde{\omega}_k^\infty(y_0)=0$. This phenomenon, first described and explained theoretically in  \cite{bouchetmorita2010}, has been called vorticity depletion at the stationary
streamlines. It has been observed numerically that the extent of the
area for which $\tilde{\omega}_{k}^{\infty}(y_{0})\simeq0$ can be
very large, up to half of the total domain, meaning that in a large
part of the domain, the shear is not the explanation for the asymptotic
decay. The formula for the vorticity (\ref{eq:orr-mechanism-voricity})
are then valid for any $y$, even for zonal jets with stationary points, provided they are stable and have no neutral modes. The formulas for the velocity and the stream function
are valid for any $y\neq y_{0}$. Exactly at the specific point $y=y_{0}$,
the damping is still algebraic with preliminary explanation given
in \cite{bouchetmorita2010}, but a complete theoretical prediction
is not yet available.\\

We have thus seen that, under the hypothesis that $\beta=0$ and that the linear operator $L_U^0$ has no normal mode, the deterministic dynamics of the eddies leads to an inviscid damping of the velocity and of the stream function. As explained in the introductory example of the one dimensional Ornstein--Uhlenbeck stochastic dynamics, this is the key ingredient that can ensure the convergence of the average momentum flux $\mathbf{E}_{U}[\langle v_{m}^{(y)}\omega_{m}\rangle ]$. We investigate this point in the following paragraph.

\subsection{Convergence of the averaged momentum flux convergence}

\begin{figure}
\figurebox{}{15pc}{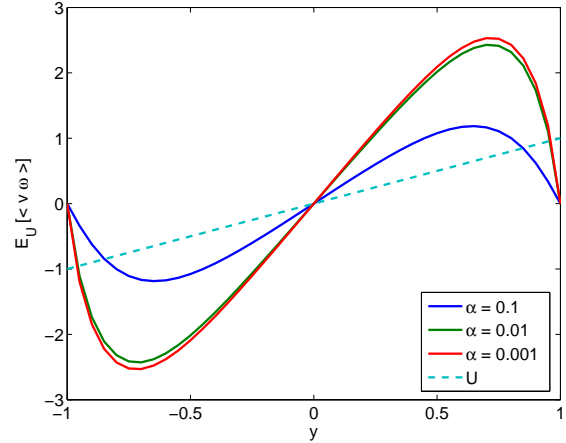}

\caption{The stationary momentum flux convergence $\mathbf{E}_{U}\left[\left\langle v_{m}^{(y)}\omega_{m}\right\rangle \right]$
in the case of a linear base profile $U(y)=y$ in a channel geometry,
with $\nu=0$ and with different values of the friction coefficient
$\alpha$. We check the convergence of this quantity to a smooth function
in the inertial limit $\alpha\to 0$. The details about the numerical computation of this quantity can be found in \cite{bouchetnardinietal2013}.\label{fig:NUM-parabolique-lima0}}
\end{figure}

Starting from the eddy equation (\ref{linear_dynamics}), a direct generalisation of equation (\ref{autocorrelation_1D}) for the average momentum flux convergence gives
\[
\mathbf{E}_{U}\left[\left\langle v_{m}^{(y)}\omega_{m}\right\rangle(y)\right]=\sum_{k>0,l}c_{kl}F_{kl}(y), \ {\rm with}
\]
\begin{equation}
F_{kl}(y)=\lim_{t\rightarrow\infty}\int_{0}^{t}\,\tilde{\omega}_{k}(y,u)\,\tilde{v}_{k}^{(y)*}(y,u)\,\mbox{d}u+\mbox{C.C.},\label{eq:hkl}
\end{equation}
where $\tilde{\omega}_{k}\mbox{e}^{ikx}$ and $\tilde{v}_{k}^{(y)}\mbox{e}^{ikx}$
are the deterministic solutions to the linearized equation $\partial_{t}\tilde{\omega}+L_{U}^{0}\tilde{\omega}=0$
with initial condition $\mbox{e}^{ikx+ily}$, $\mbox{C.C.}$ denotes the complex conjugate, and $c_{kl}$ are the Fourier components of the forcing correlation function $C_m$ (a detailed derivation is given in \cite{bouchetnardinietal2013}).\\

Using the asymptotic expressions of the deterministic fields (\ref{eq:orr-mechanism-voricity},\ref{eq:orr-mechanism-velocity-y}), we readily see that the integral in equation (\ref{eq:hkl}) converges. We have thus proven that, under the hypothesis that $\beta=0$ and that the base flow $U$ has no normal modes, the momentum flux convergence $\mathbf{E}_{U}[\langle v_{m}^{(y)}\omega_{m}\rangle ]$ converges to a finite quantity when $\alpha \to 0$. This is illustrated in figure \ref{fig:NUM-parabolique-lima0}.

In order for the time scale separation to be justified, and the kinetic equation to be valid, it is not only required that the momentum flux convergence $\mathbf{E}_{U}[\langle v_{m}^{(y)}\omega_{m}\rangle ]$ has a limit when $\alpha \rightarrow 0$, but an ergodic property
\begin{equation}
\lim_{T \rightarrow \infty} \frac{1}{T}\int_0^T \langle v_{m}^{y}\omega_{m}\rangle(y,t)\ {\rm d}t  = \mathbf{E}_{U}[\langle v_{m}^{(y)}\omega_{m}\rangle ](y),\label{eq:ergodicity}
\end{equation}
 should also be verified, and this limit should be valid uniformly with respect to $\alpha$. Again the result is not obvious as we have to count on the inviscid damping mechanism. The existence of this ergodic limit has been studied  \cite{tangarife-these}. As the development are quite technical, we just comment here the main results. First it has been proven that
\begin{equation}
\mathbf{E}_{U} \left\{ \left[ \frac{1}{T}\int_0^T \langle v_{m}^{(y)}\omega_{m}\rangle(y,t)\ {\rm d}t  - \mathbf{E}_{U}[\langle v_{m}^{(y)}\omega_{m}\rangle ](y) \right]^2 \right\} ... 
\end{equation}
\begin{equation}
 ... \underset{T \rightarrow \infty}{\sim} \frac{A(y)}{\alpha T},
\end{equation}
which seems to be a negative result about ergodicity. Indeed one can see from this result that while for finite value of $\alpha $ the large time ergodic limit is the expected one, the quadratic error diverge when $\alpha \rightarrow 0$. However it has also been proven that ergodicity occurs for the momentum flux convergence understood as distributions. This means that for any smooth test function $\phi$ one can prove that
\begin{equation}
\mathbf{E}_{U} \left\{ \left[  \int {\rm d}y \  \phi(y)\frac{1}{T}\int_0^T \langle v_{m}^{(y)}\omega_{m}\rangle(y,t)\ {\rm d}t  - ...\right. \right.
\end{equation}
\begin{equation}
... \left. \left. \int {\rm d}y  \ \phi(y) \mathbf{E}_{U}[\langle v_{m}^{(y)}\omega_{m}\rangle ](y) \right]^2 \right\} \underset{T \rightarrow \infty}{\sim} \frac{B}{T}.
\end{equation}
The quadratic error is then bounded independently on $\alpha$. Heuristically this means that the momentum flux divergence does not converge pointwise because of wild fluctuations, but as soon as those fluctuation are integrated out, the ergodic result holds. This is enough for the theory to be self consistent.

Those properties about the convergence of the statistical averages and the ergodicity property of the time averaged momentum flux convergence are essential ones for the self consistency of the theory. At a theoretical level, it means that the perturbative expansion performed in section \ref{sec:kinetic-theory} is self-consistent at the level of the momentum flux convergence. We stress that a more complete mathematical justification would also require to justify the self-consistency of the hypothesis made when we neglected the nonlinear-nonlinear eddy interactions. As a conclusion, we stress that proving the ergodicity of the momentum flux convergence is a decisive step towards a mathematical justification of the assumption (\ref{decomposition-velocity}) and of the time-scale separation. 

\subsection{Self consistent theory for the $\beta$ plane barotropic model and more complex models}

The results presented in the previous paragraphs about the ergodicity of the momentum flux convergence have been proven for the linearized Euler equation, i.e. for the case $\beta=0$. For geophysical applications, it would be very interesting to understand if these results also apply to the linearized beta-plane equation. So far, the asymptotic behavior of the linearized barotropic equation has been mostly studied in the particular case of a parabolic jet profile, such that the gradient of potential vorticity $U''(y)-\beta$ either exactly vanishes \cite{brunetwarn1990}, or is small \cite{brunethaynes1995}. In the first case, the deterministic linear dynamics can be solved explicitly, and it can be shown that an inviscid damping mechanism exits, leading to an algebraic decay of the stream function as $\tilde{\psi}_k\sim t^{-1/2}$. This decay is not fast enough to insure the convergence of the statistical average of the momentum flux convergence (\ref{eq:hkl}). In this very particular case, the present theory seems not self-consistent. However the divergence has probably a very limited spatial extension close to the jet extrema. Moreover, this case might be a very singular one, indeed the case of a small but strictly negative potential vorticity gradient \cite{brunethaynes1995} leads to a decay of the stream function as $\tilde{\psi}_k\sim t^{-3/2}$. Then the momentum flux convergence (\ref{eq:hkl}) converges for small $\alpha$, and the theory seems self-consistent.\\

The other hypothesis made to obtain the ergodicity result is that the linear operator $L_U^0$ has no normal modes, neither unstable nor neutral. While this assumption may seem restrictive at first, it is actually a generic case for the 2D Euler equation. It is indeed a classical result that shear flows without inflection points, or vortices with strictly decreasing vorticity profile are stable and have no neutral mode \cite{drazinreid1981}. The only examples of stable flows for the 2D Euler dynamics with neutral modes we are aware of, are cases with localized vorticity profile \cite{schecterdubinetal1999}.

When it comes to the linear barotropic equation, this assumption might be more restrictive. Indeed, the Rossby waves are very common neutral modes of the linearized barotropic dynamics, and are expected to exist in geophysical situations \cite{pedlosky1982}. However, we note that a mechanism of expulsion of normal modes in the presence of a background zonal jet has been revealed, and seems to hold in the atmosphere \cite{kasahara1980}. In the case where the linear dynamics would still have neutral modes, the typical time scale of propagation of the wave would be an intermediate time scale between the evolution of the jet and the evolution of the eddies. This contribution should thus be extracted from the eddies dynamics, and the effective equation of the jets dynamics would be modified accordingly. This point is currently under investigation.

\section{Comparison of theoretical predictions with numerical experiments\label{parameters}}

In section \ref{sec:S3T}, we discuss the relation of the kinetic equation described in this chapter (eq. \ref{eq:Stochastique-Lent-U-first-order}) with related approaches (the S3T-CE2-quasi-linear equations) discussed in chapters 5.2.2. and 5.1.2., respectively. In section \ref{sec:numerics} we discuss numerical experiments that confirm that the small $\alpha$ limit is actually the relevant one for the validity of the kinetic approach. Finally, in section \ref{sec:bistability} we discuss the effect of the stochastic terms appearing at higher order in our equation and explain their importance in order to determine both Gaussian and large fluctuations of the jet profile.

\subsection{Deterministic slow evolution of the zonal jets through kinetic theory and the S3T-CE2 system}
\label{sec:S3T}
As explained in section \ref{summary}, the statistical average of the momentum flux convergence appearing in the equation for the slow evolution of the zonal jet (\ref{eq:Stochastique-Lent-U-first-order}) is computed from the statistically stationary statistics of the linearized dynamics (\ref{linear_dynamics}) for $U$ held fixed. Equivalently, it can be computed as a linear transform of the stationary solution of the Lyapunov equation (\ref{eq:equation-Lyapunov}). 

The S3T-CE2 system \cite{bakasioannou2013,srinivasanyoung2011,tobiasdagonetal2011} is obtained from the quasi-linear approximation (setting to zero the non-linear eddy-eddy interaction terms in the equation on $\omega_m$)  and by moreover taking an average of the momentum flux convergence in the equation for $U$. Moreover, it is assumed that the statistical average (over the realisations of noise) coincides with a spatial average over the zonal direction $x$. The resulting equations are thus very similar to the kinetic equation (\ref{eq:Stochastique-Lent-U-first-order}). The main difference is that the jet and the correlation function of the fluctuations evolve simultaneously. In a statistically stationary state, neither the jet profile $U(y)$ nor the correlation of the fluctuations $g$ evolve. As a consequence, we can assess that our kinetic equation and the S3T-CE2 system have the same attractors. The kinetic approach is a perturbative expansion when the parameter $\alpha$ is very small.  In this limit, because of the time scale separation, the results of S3T-CE2 should coincide with the kinetic theory. We will see in next section that the direct numerical simulations of the barotropic equations are in good agreement with the S3T-CE2 equations, and thus with the kinetic theory, in this regime $\alpha \ll 1$.

A very interesting and important practical advantage of the S3T-CE2 equation is that it gives an autonomous equation that can be integrated forward in time, independently of any hypothesis. It is thus an interesting tool in order to study the dynamics, both numerically and theoretically, even when the hypothesis for the validity of the kinetic theory are not satisfied. As an example, the case of a homogeneous flow, $U=0$, that does not enter into the class of flow with no-modes considered in the kinetic approach, has been extensively studied in the S3T-CE2  framework \cite{bakasioannou2013,srinivasanyoung2011}.  One reason is that it is explicitly solvable. Those works also give a very interesting qualitative understanding of the mechanisms leading to the formation of coherent zonal flows.

However, we stress that there is no clear reason to expect the S3T-CE2 approach to give quantitatively correct results when the basic hypothesis of the kinetic theory are not verified. We recall them here: there should be a time scale separation between the evolution of the non-zonal perturbations and the slow jet dynamics (this is the case for instance if $\alpha \ll 1$), and the linear operator $L_U$ associated to the jet profile $U$ should have no normal modes. 

\subsection{Comparison of theoretical results and numerical simulations \label{sec:numerics}}
We now investigate the parameters used in numerical simulations of the S3T-CE2-quasi-linear equations. For simplicity we focus on the work by Tobias and Marston \cite{tobias2013}, but the conclusions are the same for the other works \cite{bakasioannou2013,srinivasanyoung2011}.

In this paper, it is argued that the strength of the jets is related to the value of the zonostrophy index
\begin{equation}
R_\beta = \frac{U^{1/2}\beta^{1/10}}{2^{1/2}\epsilon^{1/5}}\,,
\end{equation}
which has also been introduced in \cite{danilovgurarie2004,galperinsukorianskyetal2010}. $R_\beta$ is obtained as the ratio of the Rhines scale and of another length scale built by comparing the intensity of the forcing and of the mean gradient of potential vorticity $\beta$. It is observed that a large value of $R_\beta$ leads to a flow made of robust jets, while a small value leads to the formation of weak, meandering jets. Moreover, the comparison between CE2 calculations and direct non-linear simulations shows a very good agreement for large values of $R_\beta$, and a poor agreement for smaller values of this index.

We now compare these results with the scaling arguments and the kinetic theory presented before. First, we can note that we have the relation
\begin{equation}
\alpha_R = \frac{1}{2^{7/2}R_\beta ^{5}}\,,
\end{equation}
so that the regime $R_\beta \gg 1$, in which robust jets and good accuracy of S3T-CE2-quasi-linear approximation are found, coincides with the regime $\alpha_R \ll 1$. Let's now look more precisely at the different parameters considered in \cite{tobias2013}.

Three simulations are presented in this paper, corresponding to figures 2(a), 2(b) and 2(c), or 4(a), 4(b) and 4(c) for the comparison with the CE2 simulation. We find the following results:
\begin{itemize}
\item With the parameters of the case (a), we have $\alpha = 0.068$ and $\alpha_R = 0.0021$, which are both very small. This is in accordance with the fact that robust jets are found, and that the quasi-linear approximation is accurate.

\item With the parameters of the case (b), we find the values $\alpha = 0.068$ and $\alpha_R = 0.0029$, which are still very small. Again, this is in accordance with the fact that strong jets are found, and that the quasi-linear approximation is accurate.

\item With the parameters of the case (c), we have $\alpha = 1.45 >1$ and $\alpha_R = 0.030$, which is still quite small. The case (c) corresponds to weak and meandering jets, and to a very poor agreement between CE2 and non-linear simulation.\\
\end{itemize}

To conclude this discussion, we find that small values of $\alpha_R$ and $\alpha$ lead to the formation of strong jets, and to a very good accuracy of the kinetic equation (S3T-CE2-quasi-linear equations). This observation can also be made from the numerical simulations presented in other papers \cite{bakasioannou2013,srinivasanyoung2011}. However, the last case (c) suggests that $\alpha_R$, may have to be quite small in order for the dynamics to be in the range of validity of the quasi-linear approximation. This can also be seen in figure 6 of \cite{srinivasanyoung2011}, where the ratio of energy contained in the jets is plotted as a function of an adimensionalized friction $\mu_*$ and of an adimensionalized gradient of potential vorticity $\beta_*$. Which of $\alpha_R$ or $\alpha$ is the more relevant parameter to asses the range of validity of the quasilinear approximation? We find that strong jets, together with a good accuracy of the quasi-linear approximation, is obtained for small values of $\mu_*$, almost independently of the value of $\beta_*$. Then, it seems that the value of $\beta$ does not control the robustness of jets and the validity of the quasi-linear approximation, suggesting that $\alpha$ -- and not $\alpha_R$ that depends on $\beta$ -- is the relevant small parameter for the kinetic theory of zonal jets. It may be interesting to study further this hypothesis that $\alpha$ may be more relevant than $\alpha_R$ to asses the range of validity of the kinetic theory.

\section{Fluctuation of momentum flux convergence and bistability of Jupiter's zonal jets\label{sec:bistability}}

Taking into account the terms of order $\alpha^{2}$ allows to go further in the understanding of jets dynamics. Indeed, the first order
(\ref{eq:Stochastique-Lent-U-first-order}) only describes the relaxation of
a jet profile $U$ towards its attractor and the fluctuations due to the direct effect of the original forcing $\eta$ acting on zonal degrees of freedom. However, in most physically relevant situations, $\eta_z=\langle\eta\rangle=0$, and fluctuation of the zonal jet are due to fluctuation of the momentum flux convergence; this is the case we will consider from here on. 

At second order in $\alpha$, a new term appears: a white in time noise with spatial correlation function $\alpha \Xi_{NL}[U]$. This noise term describes both Gaussian fluctuations of the jet profile due to momentum flux convergence fluctuations during the relaxation, or Gaussian fluctuations of the jet around its attractors. It is thus an interesting correction whenever one is interested in Gaussian fluctuations of the jet.

\begin{figure}
\figurebox{18pc}{}{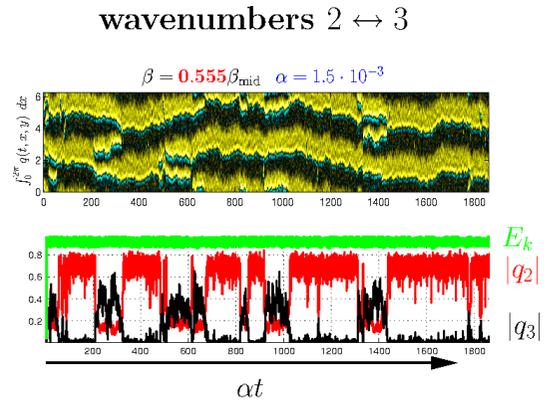}
\caption{Rare transitions between attractors, with respecpectively two and three alternating jets, for the barotropic turbulence on a beta plane. Those rare transitions are similar to the one observed on Jupiter during the period 1939-1940. The upper plot shows the zonally averaged vorticity as a function of the slow time $\alpha t$. The lower plot shows the modulus of the wave number 2 and 3 Fourier components of the zonally averaged vorticity, and the energy, as a function of time. Transitions are extremely rare: 11 transitions are observed over a time scale of  about $10^6$ turnover times (from a work with E. Simonnet).\label{fig:Transition-Stationnaire}}
\end{figure}
Another type of physical situations when fluctuations are essential is when the system is multistable because the relaxation
dynamics (\ref{eq:Stochastique-Lent-U-first-order}) has two (or more) attractors. Such a multistability seems to be relevant for Jupiter's zonal jets. Indeed during the period 1939-1940 three Jupiter's white ovals suddenly appeared, most probably following the instability and the disappearance of one of the alternating zonal jets \cite{youssef2003dynamics,rogers1995giant}. Cases of multiple attractors are also known in zonal jet dynamics for beta-plane barotropic turbulence; for example, in chapter 5.1.2 and 5.2.2, two attractors with a different number of jets are shown to emerge using the same physical parameters and different initial conditions. In Figure \ref{fig:Transition-Stationnaire} we show the first example of rare transitions between zonal jets for the beta-plane barotropic turbulence. The most amazing result is that such transitions between a state with two alternating jets, and a state with three alternating jets, are extremely rare. On the numerical simulation described in Figure \ref{fig:Transition-Stationnaire} only 11 transitions occur over a time scale of $10^6$ turnover time. Also on Jupiter such events are extremely rare; indeed since the appearance of the three white ovals about 75 years ago, no similar event has been observed. Such situations of bistability and very rare transitions are very common in geophysical, two-dimensional,
and three-dimensional turbulent flows. For instance, paths of the
Kuroshio current \cite{schmeitsdijkstra2001}, atmospheric flows \cite{weekstianetal1997},
Earth's magnetic field reversal and MHD experiments \cite{berhanumonchauxetal2007},
two--dimensional turbulence simulations and experiments \cite{sommeria1986,bouchetsimonnet2008,maassenclercxetal2003,loxleynadiga2013},
and three--dimensional flows \cite{raveletmarieetal2004}
show this kind of behaviour. 

In a bistability situation, for instance the zonal jet bistability described in Figure \ref{fig:Transition-Stationnaire}, the time the dynamics spend close to one attractor before jumping to the other is described by a Poisson statistics, determined by two transition rates. Each transition rate is the inverse of the average time spent close to one attractor before jumping to the other. The most important scientific question is to determine these two transition rates. Once those rates are known, the stationary distribution can be computed, giving access to the stationary probability to observe one or the other attractor. 

A very interesting issue is to understand if transition rates could be computed from a quasilinear approximation, and how. The key point is that such rare transitions are the consequences of very rare fluctuations of the momentum flux convergence. Such fluctuations may not be described by the spatial correlation function $\alpha \Xi_{NL}[U]$, which amount at a Gaussian approximation only. In principle, in a system with two well separated time scales, as the one described in this chapter, one can compute such rare transition by evaluating the large deviations of the fast variables. For instance in the case of the barotropic turbulence, the quasilinear approach leading to neglecting eddy-eddy nonlinearities in equation (\ref{linear_dynamics}) is in principle still valid. But then one needs to compute large deviations of time average momentum flux convergence, that is one should compute the probability to observe the quantity $\frac{1}{T}\int_0^T \langle v_{m}^{(y)}\omega_{m}\rangle(y,t)\ {\rm d}t$ to be equal to some arbitrary value $R$, rather than just the average of this quantity (giving $\mathbf{E}_{U}[\langle v_{m}^{(y)}\omega_{m}\rangle]$) or its Gaussian fluctuations (related to $\alpha \Xi_{NL}[U]$). For that purpose, the Lyapunov equation (\ref{eq:equation-Lyapunov}) is not sufficient and, one should develop a new formalism based on matrix Riccatti equations. This approach has been recently developed in \cite{tangarife-these} and is the subject of several ongoing researches. The scope of those research is to compute transition rates for Jupiter's like zonal jets.

\section{Conclusion}

In this chapter we have discussed a theory of zonal jets velocity profiles, in an inertial limit, when there is a clear separation of time scales between the rapid evolution of the turbulent non zonal part of the velocity field and the slow evolution of zonal jets. Under this hypothesis, and further assuming that the linearised equation close to the zonal jets has no unstable or neutral eigenmodes, the theory predicts the jet velocity profile and the turbulence statistics. This systematic expansion makes precise previous approaches based on quasi-linear approximations or cumulant expansion.

We foresee many further theoretical developments of this theory. For instance prediction of phase transitions, bistability, and transition rates will be studied in the future, using large deviation theory.  A more complete theoretical study of the conditions for this theory to be valid in more complex models, including layered and three dimensional quasi geostrophic models and the primitive equations should also be considered.

The applications of this theory are further discussed in the chapters 5.1.2, 5.2.2, 5.2.3, 5.2.4, and 5.2.5 of this book.

\bibliographystyle{plain}
\bibliography{bib-kinetic-theory}

\end{document}